\documentclass[11pt]{article}
\usepackage{amsmath,amssymb,amsfonts} 
\usepackage{latexsym}   
\begin{document}
\begin{titlepage}
\begin{flushright}
UTAS-PHYS-2007-22\\
December 2007\\
\end{flushright}
\begin{centering}
 
{\ }\vspace{0.5cm}
 
{\Large\bf The molecular asymmetric rigid rotor Hamiltonian
as an exactly solvable model}
\vspace{1.8cm}

\vspace{0.7cm}

P. D. Jarvis\footnote{Alexander von Humboldt Fellow} and
L. A. Yates \footnote{Australian Postgraduate Award}

\vspace{0.3cm}

{\em School of Mathematics and Physics}\\
{\em University of Tasmania, Private Bag 37}\\
{\em 7001 Hobart, Tasmania, Australia}\\
{\em E-mail: {\tt Peter.Jarvis@utas.edu.au}, {\tt Luke.Yates@utas.edu.au}}


\begin{abstract}
\noindent
Representations of the rotation group may be formulated in
second-quantised language via Schwinger's transcription of angular
momentum states onto states of an effective two-dimensional oscillator.
In the case of the molecular asymmetric rigid rotor, by projecting onto the state space of rigid body rotations,
the standard Ray Hamiltonian $H(1,\kappa,-1)$
(with asymmetry parameter $1 \ge \kappa \ge -1$), becomes a quadratic polynomial in the generators of the
associated dynamical $su(1,1)$ algebra. We point out that $H(1,\kappa,-1)$ is in fact quadratic in the Gaudin operators
arising from the quasiclassical limit of an associated $su_q(1,1)$ Yang-Baxter algebra. The general asymmetric rigid rotor Hamiltonian is thus an exactly solvable model. This fact has important implications for the structure of the spectrum, as well as for the eigenstates and correlation functions of the model.

\end{abstract}

\vspace{10pt}

\end{centering} 

\vspace{125pt}

\end{titlepage}

\setcounter{footnote}{0}
\section{Introduction}
\label{Sec1}
The analysis of rotational dynamics is central to molecular spectroscopy and has been much studied in the literature. The paradigmatic model is of course the rigid rotor, whose symmetric (prolate-oblate) limit is trivially handled in the quantum  case using the angular momentum algebra. For the asymmetric case one has an additional parameter $\kappa$, and
although, in principle, the model is easily solved for given angular momentum by a matrix diagonalisation \cite{Ray,King}, it is surprisingly rich. Classical approaches to the full, non-rigid case achieve a separation of vibrational and rotational degrees of freedom by imposing subtle geometrical constraints on the relative coordinates, which often generate expansions of considerable combinatorial complexity when accurate descriptions of couplings are developed \cite{Papousek,RileyBook,Watson2002}. Such is the fundamental importance of the general problem that new contributions continue to appear, as for example the recent work \cite{Vasquez} concerning large $N$ expansions, and \cite{Watson2007} on a semiclassical approach to expansions of the energy levels and wavefunctions near the symmetric limits $\kappa=\pm1$.

In contrast to \emph{ab initio} approaches are algebraic developments which adopt as a starting point, an appropriate Lie algebra of operators with enough structure to capture important features of the molecular Hamiltonian (such as interactions which dominate near level degeneracies), but where results from representation theory can be a valuable organising framework. An early study of potential dynamical symmetry groups for the rotor problem is \cite{Bohm}; unitary dynamical symmetries have been extensively used in  \cite{IachelloBook}. In the end such treatments, as well as the above-mentioned formal expansions from classical dynamics, tend to become semi-empirical studies of appropriate parametrisations in an attempt to reproduce specific aspects of observed data. 

A problem equivalent to the rigid rotor model was studied some time ago by Patera and Winternitz \cite{Winternitz} in the context of rotation group representation basis labelling. In that work, the system was shown to be separable in elliptic coordinates on the sphere, and the eigenfunctions and eigenvalues turned out to be related to certain Lam\'{e} polynomials, with the asymmetry $\kappa$ being essentially the modulus for the Jacobian elliptic functions involved in the coordinate system. Thus, to the extent that the relevant Lam\'{e} polynomials are available and tabulated, the problem can be regarded as completely solved in principle by this analytical method, in a similar way that it can be said to be solved on the algebraic side, by a straightforward matrix diagonalisation. 

The present work is intermediate between the classical dynamics-inspired works, and the use of algebraic models as dynamical symmetries, while complementing the analytical approach. It extends the well-known, textbook fact that the symmetric (oblate-prolate) limit of the rigid rotor is diagonalisable in standard angular momentum bases, to the class of molecular asymmetric rotor Hamiltonians, which are shown to be exactly solvable in the sense of Bethe \emph{Ansatz} techniques (see below).
 
In \S \ref{Sec2} below we review some transcriptions of the rotor problem using well-known algebraic tools, chiefly the model of Schwinger \cite{Schwinger} which gives a realisation of the $su(2)$ angular momentum algebra in the second-quantised language of modes of a two-dimensional oscillator. The relation to the algebra of the non-compact symmetry group $su(1,1)$ is also noted \cite{BiedenharnHolman1966}. In \S \ref{Sec3}, the symmetric and asymmetric rigid rotor Hamiltonians (in the Ray \cite{Ray} parametrisation $H(1,\kappa,-1)$ ) are compared with the class of so-called Gaudin operators \cite{Gaudin}. These are terms in the expansion of transfer operators coming from a specific realisation of a certain Yang-Baxter algebra in the so-called quasi-classical limit. To the extent that they are coefficients in $u$ expansions of formal operators $t(u)$ which mutually commute, $[t(u),t(v)]=0$ for all $u,v$, the Gaudin operators, or any polynomials thereof, are simultaneously diagonalisable via Bethe \emph{Ansatz} techniques in the tradition of exactly-solvable models in statistical and condensed matter physics \cite{FaddeevEtAl,Faddeev,Sklyanin}. The symmetric rotor (as might be expected) can trivially be identified with such transfer matrix coefficients for the two-dimensional Yang-Baxter algebra with rational $R$-matrix (the $su(1,1)$ case), whereas the extension to the trigonometric $su_q(1,1)$ case yields, in the quasi-classical Gaudin limit, the asymmetric rigid rotor Hamiltonian. The deformation $q=e^{i\gamma}$ provides the additional asymmetry parameter, the identification being essentially $\cos \gamma = \kappa$
(after rescaling $\gamma \rightarrow \gamma \eta, u \rightarrow u/\eta$ and taking the limit $\eta \rightarrow 0$).

The conclusion of our investigation is therefore that the molecular asymmetric rigid rotor does indeed belong to the exactly solvable class.  This fact has important implications for the structure of the spectrum, as well as for the eigenstates and correlation functions of the model. Our work also establishes hitherto unnoticed connections between the analytical approach via special functions, and integrable models of this type. Further discussion and conclusions are given in  \S \ref{Sec4} below. 

\section{Oscillator formalism}
\label{Sec2}
The energy operator for the asymmetric rigid rotor when written in body-fixed coordinates adapted to the principal moments of inertia of the rigid body takes the form
\begin{align}
{H} =& \, \frac{J_a^2}{2I_a} + \frac{J_b^2}{2I_b} +\frac{J_c^2}{2I_c}, \nonumber
\end{align}
where $J_a, J_b, J_c$ are the generators of rotations about the body-fixed axes, and $I_a, I_b , I_c$ are the principal moments of inertia. For future reference we define 
$a = (2I_a)^{-1}$, $b = (2I_b)^{-1}$, $c=(2I_c)^{-1}$, so that the Hamiltonian becomes
\begin{align}
H(a,b,c) & = \, a J_a^2 + b J_b^2 + cJ_c^2.
\end{align}
The standard way of writing the rigid rotor Hamiltonian is via the so-called Ray \cite{Ray} Hamiltonian $H(1,\kappa,-1)$ defined by deriving the simple relation
\begin{align}
H(a,b,c) = & \, \textstyle{\frac 12}(a-c)H(1,\kappa,-1) + \textstyle{\frac 12 }(a+c)H(1,1,1), \label{eq:KingHamDef} \\
\kappa = & \, \frac{2b-a-c}{a-c}, \quad 1 \ge \kappa \ge -1 \nonumber
\end{align}
where it is assumed that $a \ge b \ge c$, and \emph{a fortiori}, $1 \ge \kappa \ge -1$, and $H(1,1,1)$ plays the role of the Casimir operator $C(su(2))$. In this case the oblate and prolate cases are $\kappa = +1$ and $\kappa = -1$, respectively. 

States of the system are spanned by the set $|{\mathcal J},{\mathcal M} ; j, m \rangle$, where 
${\mathcal J}({\mathcal J}+1)$ is the eigenvalue of the square of the total (centre-of-mass) angular momentum operator, with ${\mathcal J}$ the total angular momentum quantum number, and ${\mathcal M}$ its third component; we have 
as usual ${\mathcal M} = -{\mathcal J}, -{\mathcal J}+1, \cdots, {\mathcal J}-1, {\mathcal J}$. $j \equiv {\mathcal J}$ is similarly the angular momentum quantum number for body-fixed rotations. Thus each eigenstate of $H(a,b,c)$ is $2{\mathcal J}+1$-fold degenerate, corresponding to the values of ${\mathcal M}$, and the Hamiltonian must be diagonalised, for fixed ${\mathcal J}=j$, 
identically for each ${\mathcal M}$, only in the $(2j+1)\times (2j+1)$ subspace of states of differing $m$, $ m = -j, -j+1, \cdots , j-1,j$. We henceforth drop explicit reference to the ${\mathcal J},{\mathcal M}$ labels and refer to the basis simply as $|j,m \rangle$ -- in other words, we are exploiting conservation of angular momentum, and referring the solution of the complete rigid rotor problem on the space 
 $\oplus_{{\mathcal J}\equiv j=0}^\infty \, H_{\mathcal J} \otimes H_j$, to the diagonalisation of an \emph{effective} rigid rotor Hamiltonian on the model space $\oplus_{j=0}^\infty H_j\,$. The \emph{energy eigenstates} can be denoted $||j, \tau \rangle\!\rangle$, where $\tau$ is a pseudo-magnetic quantum number labelling states of increasing energy and running over $2j+1$ values $-j, -j+1, \cdots, j$ (see \cite{Ray,King}). In this notation it is known \cite{Ray} that the corresponding energy eigenvalues $E_\kappa({j,\tau})$ satisfy the reflection property, $E_\kappa({j,\tau}) = E_{-\kappa}({j,-\tau})$, so that the rigid rotor problem essentially amounts to diagonalising the one-parameter family $H(1,\kappa,-1)$ in the restricted interval $0\le \kappa \le 1$. 
 
In the formulation of Schwinger \cite{Schwinger}, the $su(2)$ algebra of angular momentum can be expressed in terms of the raising and lowering modes of a set of two independent oscillators (which can be thought of as associated to an auxiliary two-dimensional space). Calling these mode operators $\widehat{a}, \widehat{a}^\dagger$, $\widehat{b}, \widehat{b}^\dagger$, 
the angular momentum generators are
\begin{align}
J_x = & \, -\textstyle{\frac 12} i (\widehat{a}^\dagger \widehat{b} - \widehat{b}^\dagger \widehat{a}), \qquad J_y = \textstyle{\frac 12} (\widehat{a}^\dagger \widehat{b} + \widehat{b}^\dagger \widehat{a}),
\qquad J_z = \textstyle{\frac 12}(\widehat{a}^\dagger \widehat{a} - \widehat{b}^\dagger \widehat{b}) 
\end{align}
which can be easily checked to fulfil the required $su(2)$ Lie algebra relations,
\begin{align}
{[}J_x, J_y {]} = i J_z, \quad \mbox{and cyclically}, \nonumber
\end{align}
provided the oscillator raising and lowering operators fulfil the usual relations
\begin{align}
{[} \widehat{a}, \widehat{a}^\dagger {]} = & \,  1, \qquad {[} \widehat{a}, \widehat{a}  {]} = {[} \widehat{a}^\dagger, \widehat{a}^\dagger {]} = 0, \nonumber \\
{[} \widehat{b}, \widehat{b}^\dagger {]} = & \,  1, \qquad {[} \widehat{b}, \widehat{b}  {]} = {[} \widehat{b}^\dagger, \widehat{b}^\dagger {]} = 0, \nonumber \\
\mbox{with of course} \nonumber \\
{[} \widehat{a}, \widehat{b}  {]} = & \, {[} \widehat{a}, \widehat{b}^\dagger {]} =  {[} \widehat{a}^\dagger, \widehat{b} {]} = {[} \widehat{a}^\dagger, \widehat{b}^\dagger {]} =0.
\end{align}

The advantage of the Schwinger transcription is that angular momentum eigenstates $|j,m\rangle$ as above can explicitly be written down in terms of Fock mode states created by monomials in the $\widehat{a}^\dagger$, $\widehat{b}^\dagger$ operators acting on the vacuum state $|0,0\rangle$ defined by
\begin{align}
& \, \widehat{a}|0,0\rangle = \widehat{b} |0,0\rangle = 0; \nonumber \\
| N_A, N_B \rangle = & \, \frac{(\widehat{a}^\dagger)^{N_A}}{\sqrt{N_A!} } \frac{(\widehat{b}^\dagger)^{N_B}}{\sqrt{N_B !}}|0,0 \rangle, \quad N_A, N_B = 0,1,2,\cdots, \nonumber  \\
\mbox{with} \qquad  \widehat{N}_A | N_A, N_B \rangle = & \, N_A | N_A, N_B \rangle, \qquad 
\widehat{N}_B | N_A, N_B \rangle = N_B | N_A, N_B \rangle, \qquad \mbox{where} \nonumber \\
\widehat{N}_A := & \,  \widehat{a}^\dagger \widehat{a}, \qquad \widehat{N}_B := \widehat{b}^\dagger \widehat{b}. 
\end{align}
Moreover, it is easily seen that the angular momentum quantum number itself is intimately related to the total  number operator for oscillator modes
\begin{align}
\widehat{N} = & \, \widehat{N}_A + \widehat{N}_B.
\end{align} 
Explicitly, an easy calculation gives
\begin{align}
C(su(2)) = & \, J_x^2+J_y^2+J_z^2 = \textstyle{\frac 12}\widehat{N}(\textstyle{\frac 12} \widehat{N}+1),
\end{align}
so that the angular momentum operator is $\widehat{j} \equiv \frac 12 \widehat{N}$ with eigenvalue $j = \frac 12 N = \frac 12 (N_A+N_B)$ on the above states. Thus, in terms of the $|j,m \rangle$ states we have in fact
\begin{align}
|N_A, N_B \rangle \equiv & \, \big| \, j= \textstyle{\frac 12}(N_A+N_B), \, m= \textstyle{\frac 12}(N_A-N_B) \, \big\rangle . 
\end{align}
Any operator acting on states of the rotor system can now be expressed in terms of monomials in the creation and and annihilation operators in the equivalent number basis, the asymmetric rotor Hamiltonian itself being a case in point.

Before proceeding, it is useful to make a different operator transcription, this time from the Schwinger angular momentum generators to the generators of a closely related algebra, also realised in terms of the oscillator modes, namely the Lie algebra $su(1,1)$ \cite{BiedenharnHolman1966}.
The latter algebra has generators $K_+,K_-,K_0$ and nonzero commutation relations
\begin{align}
{[}K_+, K_-{]} = & \, -2 K_0, \qquad {[}K_0, K_\pm {]} = \pm K_\pm, \nonumber
\end{align}
which are fulfilled by the following operators acting on the $A$, $B$ Fock spaces separately:
\begin{align}
K_+^A = \textstyle{\frac 12} \widehat{a}^\dagger \widehat{a}^\dagger,  \qquad K_-^A = \textstyle{\frac 12} \widehat{a}\widehat{a}, \qquad K_0^A = \textstyle{\frac 12} \widehat{a}^\dagger \widehat{a} + \textstyle{\frac 12}, \nonumber \\
K_+^B = \textstyle{\frac 12} \widehat{b}^\dagger \widehat{b}^\dagger,  \qquad K_-^B = \textstyle{\frac 12} \widehat{b}\widehat{b}, \qquad K_0^B = \textstyle{\frac 12} \widehat{b}^\dagger \widehat{b} + \textstyle{\frac 12}, \end{align}
as can easily be seen from the fundamental commutation relations of the oscillator modes.
The Casimir operator
\begin{align}
2C(su(1,1)) = & \, K_+K_- + K_-K_+ - 2 K_0^2 \nonumber
\end{align}
for the representations of $su(1,1)$ on each oscillator space has fixed eigenvalue $C^{A} = -\textstyle{\frac{3}{16}}= C^{B}$ (corresponding to the common eigenvalue $k(k+1)$ for the direct sum of two irreducible representations with spin quantum numbers $k = -\textstyle{\frac 14}, -\textstyle{\frac 34}$ in each case). By contrast, the total $su(1,1)$ algebra with generators $K_\pm = K_\pm^A + K_\pm^B$, $K_0 = K_0^A+K_0^B$ has Casimir
\begin{align}
C^{A+B}(su(1,1)) =& \, \textstyle{\frac 14}(N_A-N_B)^2 -\textstyle{\frac 14}
\end{align}
(corresponding to spin quantum number $k^{A+B} = -\textstyle{\frac 12}|N_A-N_B|-\textstyle{\frac 12}$).

From the above definitions we finally have general second-quantised forms for the asymmetric rigid rotor Hamiltonian, once an identification between $J_x,J_y,J_z$ and $J_a,J_b,J_c$ is made, which is to say an identification between the $x$,$y$,$z$ directions and the (oriented) principal body axes corresponding to $a$,$b$,$c$. The canonical choice, in view of the distinguished role of the $b$ axis (in the Ray parametrisation) \emph{vis a vis} the $z$ axis (with respect to the standard choice of deformation of the $su(2)$ Lie algebra, as will be seen below), turns out to be $J_x\equiv J_a$, $J_y\equiv -J_c$, and $J_z\equiv J_b$, giving straightforwardly
\begin{align}
H(a,b,c) =& \, \textstyle{\frac 12}(a-c)\big(J_a^2 - J_c^2\big) +\textstyle{\frac 12} (a+c)\big(J_a^2 + J_c^2\big) + b J_b^2
\nonumber \\
= & \, (a-c)\big(K_+^AK_-^B + K_+^AK_-^B ) + (a+c)\big( 2K_0^AK_0^B - \textstyle{\frac 18} \big) + b \big(K_0^A-K_0^B\big)^2 
\nonumber
\\
\equiv & \, (a-c)\big((K_+^AK_-^B + K_+^AK_-^B)  - 2\kappa K_0^AK_0^B\big) + b \big(K_0^A+K_0^B\big)^2 
\label{eq:AsymmRotorKForm}
\end{align}
(up to a constant), where the last rearrangement also presages the discussion to follow.

\section{Exactly solvable rotor Hamiltonians from $su(1,1)$ and $su_q(1,1)$ Gaudin operators. }
\label{Sec3}
In the theory of integrable quantum Hamiltonians \cite{FaddeevEtAl,Faddeev,Sklyanin}, a family of simultaneously diagonalisable operators may be generated from a set of algebraic equations called the $RTT$ relations. Exactly solvable Hamiltonians may be constructed as polynomials in these operators. The structure of the relations is determined by a given $R$-matrix, a (numerical) solution to the Yang-Baxter equation, and the $RTT$ relations themselves generate the Yang-Baxter algebra. Here we consider for the two dimensional case the so called rational solution, its generalisation to the trigonometric solution, which for appropriate real forms generate the algebras $Y(su(1,1))$ and $Y(su_q(1,1))$ respectively, and finally the so-called quasiclassical limit of the latter. Details of the constructions are given in the reviews cited; see for example also \cite{Linksetal}.
\subsection{Rational $R$-matrix: $su(1,1)$ transfer operator and the symmetric rotor.}
The rational solution of the Yang-Baxter equation in the two-dimensional case is the $4\times 4$ numerical $R$ matrix,
\begin{align}
R(u) = & \, \frac{1}{u}\left(\begin{array}{cccc} u+\eta &0 &0 &0 \\
                             0 & u & \eta & 0 \\
                               0 & \eta & u & 0 \\
                               0 & 0 & 0 & u+\eta \end{array} \right), \label{eq:RationalR}
\end{align}
which generates the Yang-Baxter algebra via the $RTT$ relation using the $2 \times 2$ $L$ operator
\begin{align}
L^J(u) = & \frac 1 u \left( \begin{array}{cc} u + \eta J_0 & \eta J_- \\
                            \eta J_+ & u - \eta J_0 \end{array} \right), \nonumber \\ 
L^K(u) = & \frac 1 u \left( \begin{array}{cc} u + \eta K_0 & \eta K_- \\
                           - \eta K_+ & u - \eta K_0 \end{array} \right), \label{eq:RationalL^K}
\end{align}
where in the compact case $J_\pm, J_0$ are generators of the $su(2)$ Lie algebra, and in the noncompact case the $K_0, K_\pm$ generate $su(1,1)$. In either case the monodromy operators $T(u)$ are obtained as matrix products; for the present case we have simply
\begin{align}
 T(u) = & \, \left(\begin{array}{cc} e^{\delta\eta} & 0 \\ 0 & e^{-\delta\eta} \end{array} \right) \cdot L^{A}(u-\varepsilon_A)
\cdot L^{B}(u-\varepsilon_B) \nonumber
\end{align}
for a tensor product of just two copies $A,B$ of the generators, where $\delta, \varepsilon_A, \varepsilon_B$ are parameters;  the above expressions are explicitly functions of $u/\eta$, in order to facilitate expansion of the operators about  $u \rightarrow \infty$ via the limit $\eta \rightarrow 0$ of the parameter $\eta$.

From (\ref{eq:RationalR}), (\ref{eq:RationalL^K}) we have for the transfer operator, the trace $t(u) = tr(T(u))$,
\begin{align}
t(u) = & \, \big[ 2 + \eta^2 \delta^2 \big]+
\eta^2\delta \big[ \frac{{K}_0^A}{(u- \varepsilon_A)} + \frac{{K}_0^B}{ (u- \varepsilon_B)}\big]
\nonumber \\
 & \, + \frac{\eta^2}{(u- \varepsilon_A)(u- \varepsilon_B)}\big[ 2 {K}_0^A{K}_0^B - \big({K}_+^A{K}_-^B+ {K}_-^A{K}_+^B\big) \big] + O(\eta^3).\nonumber
\end{align}
We define
\begin{align}
\tau_A = & \, \lim_{u \rightarrow \varepsilon_A} 
             \frac{(u-\varepsilon_A)(u-\varepsilon_B)}{\eta^2}t(u), \nonumber \\
\tau_B = & \, \lim_{u \rightarrow \varepsilon_B} 
             \frac{(u-\varepsilon_A)(u-\varepsilon_B)}{\eta^2}t(u), \label{eq:TauDefn} 
\end{align}
obtaining
\begin{align}
\tau_A = & \, \delta(\varepsilon_A-\varepsilon_B)K_0^A + 
              2K_0^AK_0^B - (K_+^AK_-^B+ K_-^AK_+^B), \nonumber \\
\tau_B = & \, -\delta(\varepsilon_A-\varepsilon_B)K_0^B + 
              2K_0^AK_0^B - (K_+^AK_-^B+ K_-^AK_+^B).  \label{eq:TauExpn} 
\end{align}
In comparison with the asymmetric rotor Hamiltonian, setting $\Delta \varepsilon \equiv (\varepsilon_A-\varepsilon_B)$ as the (arbitrary) spectral parameter shift, it is clear that the combinations $\frac 12(\tau_A+\tau_B)$ and  
$(\tau_A-\tau_B)/(2\delta\Delta \varepsilon)$, will produce (after setting the parameter $\delta$ to $0$) the independent bracketed terms 
in (\ref{eq:AsymmRotorKForm}), with however the restriction for the first, Casimir-type term, that $\kappa\equiv +1$; that is, the oblate $a=b>c$ case. (The prolate case follows with use of the general reflection property mentioned above).
The conclusion of this reformulation, as might be expected for this rather trivial, non-deformed case, is that the rigid symmetric (prolate/oblate) rotor is recovered as an instance of an exactly-solvable model associated with the $su(1,1)$ Yang-Baxter algebra.

\subsection{Trigonometric $R$-matrix: $su_q(1,1)$ transfer operator, quasiclassical limit, and the asymmetric rotor.}
To go further using the same algebraic template, an additional parameter is needed. This is provided by the trigonometric  generalisation of the rational solution of the Yang-Baxter equation. The trigonometric $R$-matrix which generalises (\ref{eq:RationalR}), is the numerical $4\times 4$ matrix \cite{Faddeev} 
\begin{align}
R(u;\gamma) = & \, \frac{1}{{\sin(\gamma u)}} \left(\begin{array}{cccc} {\sin(\gamma(u+\eta))}  &0&0&0 \\
                              0& {\sin(\gamma u)} &  {\sin(\gamma\eta)}  &0 \\
                              0 & \,{\sin(\gamma\eta)}\,  & \, {\sin(\gamma u)}\, &0 \\
                               0&0&0& \sin(\gamma(u+\eta)) \end{array} \right), \nonumber
\end{align}
which now is a function of the (scaled) quantities $u/\eta$ and the deformation $q = e^{i\gamma \eta}$.
The corresponding $L$-operator is 
\begin{align}
L^J(u;\gamma) = & \, \frac{1}{\sin(\gamma u)}\left(\begin{array}{cc} \sin(\gamma(u+\eta\widetilde{J}_0)) & \sin(\gamma\eta) \widetilde{J}_-  \\
                               \sin(\gamma\eta) \widetilde{J}_+ &  \sin(\gamma(u-\eta\widetilde{J}_0)) \end{array} \right), \nonumber
\end{align}
where $ \widetilde{J}_0, \widetilde{J}_\pm$ satisfy the $su_q(2)$ algebra
\begin{align}
{[} \widetilde{J}_0, \widetilde{J}_\pm {]} = & \, \pm  \widetilde{J}_\pm, \qquad 
{[} \widetilde{J}_+, \widetilde{J}_- {]} = \frac{\sin(2\gamma \widetilde{J}_0)}{\sin\gamma}; \nonumber
\end{align}
or alternatively
\begin{align}
L^K(u;\gamma) = & \, \frac{1}{\sin(\gamma u)}
\left(\begin{array}{cc}  \sin(\gamma(u+\eta\widetilde{K}_0)) & 
                                                                                 {\sin(\gamma \eta)} \widetilde{K}_-  \\
                              - {\sin(\gamma \eta)} \widetilde{K}_+ &  
                              {\sin(\gamma(u-\eta\widetilde{K}_0))}  \end{array} \right), \nonumber
\end{align}
where $ \widetilde{K}_0, \widetilde{K}_\pm$ satisfy the $su_q(1,1)$ algebra
\begin{align}
{[} \widetilde{K}_0, \widetilde{K}_\pm {]} = & \, \pm  \widetilde{K}_\pm, \qquad 
{[} \widetilde{K}_+, \widetilde{K}_- {]} = -\frac{\sin(2\gamma \widetilde{K}_0)}{\sin\gamma}. \label{eq:SUq11Alg}
\end{align}
We note the expansion,
\begin{align}
L(u;\gamma) = & \, \left(\begin{array}{cc}1\, &\, 0 \\ 0\, &\, 1 \end{array}\right) +  \eta\frac{\gamma}{\sin(\gamma u)}\left(\begin{array}{cc}\cos(\gamma u) K_0 & K_- \\
                                  - K_+ & -\cos(\gamma u) K_0 \end{array}\right) - 
        \textstyle{\frac 12} \eta^2 \gamma^2 (K_0){}^2
        \left(\begin{array}{cc}1\, &\, 0 \\ 0\, &\, 1 \end{array}\right) + O(\eta^3) \nonumber
\end{align}
so that in addition to the standard asymptotic limits of $L$ (and similarly $R$) as unit matrices for $u \rightarrow \infty$ (or $\eta \rightarrow 0$), we see that for $\gamma \rightarrow 0$ ($q \rightarrow 1)$ the trigonometric solutions revert to the ordinary rational case:
\begin{align}
\lim_{\gamma\rightarrow 0}L(u;\gamma) = 
                              & \, \frac{1}{u}\left(\begin{array}{cc} u+\eta K_0 & \eta K_- \\
                                              -\eta K_+ & u-\eta K_0 \end{array} \right).
\end{align}
Now consider the following monodromy operator $T(u)$ defined as a $2 \times 2$ matrix product over the enveloping algebra of the $\widetilde{K}{}^A$ and $\widetilde{K}{}^B$ generators,
\begin{align}
 T(u) = & \, \left(\begin{array}{cc} e^{\delta\eta} & 0 \\ 0 & e^{-\delta\eta} \end{array} \right) \cdot L^{A}(u-\varepsilon_A)
\cdot L^{B}(u-\varepsilon_B) \nonumber
\end{align}
(suppressing the $\gamma$ label in the notation). For present purposes it is sufficient to record merely the transfer matrix $t(u) := tr(T(u))$ in the scaling limit $\eta\rightarrow 0$,
\begin{align}
t(u) = & \, \big[2 + \eta^2 \delta^2\big]+
2\eta^2 \gamma \delta \big[\cot(\gamma(u- \varepsilon_A)){K}_0^A +  \cot(\gamma(u- \varepsilon_B)){K}_0^B \big]  \nonumber \\
& \, +{\eta^2 \gamma^2}
\big[2\cot(\gamma(u- \varepsilon_A))\cot(\gamma(u- \varepsilon_B)){K}_0^A{K}_0^B-\csc(\gamma(u- \varepsilon_A))\csc(\gamma(u- \varepsilon_B))({K}_+^A{K}_-^B+ {K}_-^A{K}_+^B) \big]
\nonumber \\
& \, -\gamma^2  \big[(K_0^A){}^2 + (K_0^B){}^2\big]
+ O(\eta^3); \nonumber
\end{align}
note that we recover the standard $su(1,1)$ Lie algebra from (\ref{eq:SUq11Alg}) as $q\rightarrow 1$, or just $\eta \rightarrow 0$. We define
\begin{align}
\tau_A = & \, \lim_{u \rightarrow \varepsilon_A} 
             \frac{\sin(\gamma(u-\varepsilon_A))\sin(\gamma(u-\varepsilon_B))}{\eta^2}t(u), \nonumber \\
\tau_B = & \, \lim_{u \rightarrow \varepsilon_B} 
             \frac{\sin(\gamma(u-\varepsilon_A))\sin(\gamma(u-\varepsilon_B))}{\eta^2}t(u), \label{eq:TauQDefn} 
\end{align}
obtaining
\begin{align}
\tau_A = & \, 2\delta \gamma\sin(\gamma(\varepsilon_A\!-\!\varepsilon_B))K_0^A + 
              2\gamma^2\cos(\gamma(\varepsilon_A\!-\!\varepsilon_B))K_0^AK_0^B -
              \gamma^2(K_+^AK_-^B\!+ \!K_-^AK_+^B), \nonumber \\
\tau_B = &  -\!2\delta \gamma\sin(\gamma(\varepsilon_A\!-\!\varepsilon_B))K_0^B + 
              2\gamma^2\cos(\gamma(\varepsilon_B\!-\!\varepsilon_A))K_0^AK_0^B -
              \gamma^2(K_+^AK_-^B\!+ \! K_-^AK_+^B). \label{eq:TauQExpn} 
\end{align}
Finally we can compare appropriate combinations of $\tau_A,\tau_B$ with the desired asymmetric rigid rotor in second-quantised form. Again $\frac 12(\tau_A+\tau_B)/\gamma^2$ and $(\tau_A-\tau_B)/(2\gamma\delta\Delta\varepsilon)$ provide the correct material (after setting the parameter $\delta$ to $0$) to produce the two bracketed terms in (\ref{eq:AsymmRotorKForm}); in contrast to the situation for (\ref{eq:TauExpn}) however, there is now no longer a restriction on the relative weights within the first, Casimir-type term, and we identify
\begin{align}
 \kappa := & \, \cos(\gamma\Delta\varepsilon).
\label{eq:AsymmetryGammaKappa}
 \end{align}
Thus, the molecular asymmetric rigid rotor Hamiltonian can be recovered as a polynomial in the independent Gaudin operators $\tau_A, \tau_B$. The deformation parameter $\gamma$, in the quasiclassical scaling limit, is thereby given via 
(\ref{eq:AsymmetryGammaKappa}) in terms of the Ray asymmetry parameter $\kappa$, in units defined by the spectral parameter shift $\Delta \varepsilon$ which plays the role of an (arbitrary) auxiliary parameter in the reformulation.                        

\section{Discussion}
\label{Sec4}
In this note we have established that the classic molecular asymmetric rigid rotor Hamiltonian belongs to the exactly-solvable class of models, in the tradition of statistical and many body condensed matter physics. As pointed out in the introduction, the problem is also `solvable'  in principle, in the usual sense, either algebraically via matrix diagonalisation, or analytically via special functions as solution families of differential equations deriving from Schr\"{o}dinger's equation. These standard methods 
of course increase in complexity with increasing angular momentum, and do not have much to say about the general structure of the eigenfunctions and eigenvalues. 

By contrast, our reformulation of the model as an exactly solvable system has important implications for the general structure of the solutions. For example, the algebraic Bethe \emph{Ansatz} implies that the eigenfunctions (for arbitrary angular momentum) can be given a direct product form, with parameters satisfying a set of algebraic equations (in the quasiclassical Gaudin limit); the energy eigenvalues are in turn able to be written down explicitly as polynomials in these \emph{Ansatz} parameters. Moreover, the Bethe \emph{Ansatz} solutions also open the way to exact evaluation of correlation functions, and hence to the physical properties of the system. At the same time, our work implies that there are close and hitherto unnoticed links between the manipulations involved in the special function methods applicable in this case (such as continued fraction expansions for the Lam\'{e} functions, as discussed in \cite{Winternitz}), and exactly solvable models.

At this point our analysis is on a par with analogous studies pointing out the relevance of exactly-solvable models in a range of problems in many-body and condensed-matter contexts (see for example the review \cite{Linksetal} and the paper \cite{Ortiz} on Gaudin algebras). One point to note is that, whereas the $N$-body systems often involve spin chains with a number of systems with \emph{finite} state spaces, (and for example, $su(2)$ symmetry algebras), in our second-quantised reformulation, the `chain' consists of just two systems $A$ and $B$, but with each having an \emph{infinite} number of states. Moreover, the relevant symmetry algebra is the non-compact $su(1,1)$ Lie algebra, in contrast to what might have been expected, given the obvious analogy between the asymmetric rigid rotor and anisotropic `XXZ' or `XYZ' type spin chains where deformations of $su(2)$ are involved. For further details of the notation and main results needed from exactly solvable models in many body quantum theory, we refer the reader the literature.

One final comment should be made regarding the status of the work in the context of group representation theory, which was the focus of the paper \cite{Winternitz}. In that paper the central question was the use of a non-subgroup labelling scheme for representations of the rotation group $SO(3)\cong SU(2)$. Our transcription shows that the choice needed for the asymmetric rigid rotor, essentially to replace the usual Cartan generator of rotations about a fixed axis (corresponding to the magnetic quantum number) by a quadratic, but non-Casimir, in the generators, fortuitously turns out to belong technically to what in the general case is a so-called Bethe subalgebra of the associated Yang-Baxter algebra -- in the two-dimensional case, just a function of the independent pieces of the transfer operator as we have seen. However, the situation becomes less fortuitous when compared with the analogous labelling problem for the three dimensional unitary group $SU(3)$ when, in many physical applications, the natural angular momentum embedding $SU(3) \supset SO(3)$ requires the use of 
the orbital angular momentum and orbital magnetic quantum numbers as usual, but also, an additional so-called \emph{missing label} needed to remove degeneracy in the reduction of representations of $SU(3)$ to their angular momentum constituents. It has been shown in \cite{JarvisZhang} that, of the known non-subgroup, $SO(3)$  invariant operators in the enveloping algebra of $SU(3)$, there is a unique combination which belongs to the Bethe (maximal abelian) subalgebra of the relevant Yang-Baxter algebra for this case. Although the spectra of the candidate operators in this $SU(3)\supset SO(3)$ case are known to be irrational, Bethe \emph{Ansatz} considerations would suggest that the eigenvalues of the distinguished combination are at least algebraic, and indeed, an operator derived from a Bethe subalgebra would appear to be best choice for such non-subgroup labelling situations -- just as in the present problem of the asymmetric rigid rotor.

The rigid rotor as a dynamical system is a seminal problem and has been the subject of innumerable papers. Its quantum version, especially the symmetric case, is a textbook example, while the asymmetric case is surprisingly rich, and is a basic starting point for molecular studies. It is of great interest that a range of new  tools is available for its analysis via the techniques of exactly-solvable models. We defer the detailed investigation of such further implications of our transcription to future work. 
\section*{Acknowledgements}
The authors wish to thank Mark Riley and Peter Schwerdtfeger for useful discussions during the completion of this work.
We are also grateful for hospitality and timely advice during a visit to the Mathematical Physics group, University of Queensland. We especially thank Jon Links and Tony Bracken for drawing our attention to several references, and explaining aspects of exactly solvable models to us. 
%
%


\label{lastpage}

\end{document}